\documentclass[12pt,a4paper]{article}

\usepackage{float}

\usepackage{tikz}
\usetikzlibrary{arrows}
\tikzstyle{block}=[draw opacity=0.7,line width=1cm]
\usepackage{mdframed}
\usepackage{xcolor}
\usepackage{algpseudocode}
\usepackage{amsmath, amsthm, amsfonts, amssymb}
\usepackage[utf8]{inputenc}
\usepackage{algorithm}
\usepackage{placeins}
\usepackage{rotating}
\usepackage{booktabs}
\usepackage{multirow} 
\usepackage{graphicx}
\usepackage[font=small,labelfont=bf,tableposition=top]{caption}
\usepackage{nameref}
\usepackage{subcaption}  
\usepackage{caption}
\usepackage{xfrac}
\usepackage{relsize}
\usepackage{mathrsfs}
\usepackage{lineno}

\usepackage{enumitem}
\usepackage{subcaption}
\usepackage{blindtext}
\usepackage{hyperref}
\usepackage{subcaption}

\usepackage[text={16cm,23cm},centering]{geometry}

\usepackage{amsthm,amssymb,amsmath}

\usepackage[bottom]{footmisc}
\usepackage{mathtools}

\usepackage[utf8]{inputenc}

\usepackage{setspace}
\usepackage{amsmath,amsfonts,amssymb}
\usepackage{algorithm,algpseudocode}
\usepackage{setspace}
\usepackage{bibentry}
\usepackage[bottom]{footmisc}

\usepackage{charter}

\usepackage{booktabs,caption,multicol,hhline,tikz,multirow,array}
\usetikzlibrary{arrows.meta}
\usepackage{colortbl}
\usepackage{arydshln}
\usepackage{url}
\usepackage{subcaption}

\usepackage{graphicx, multirow, array,amsmath}
\usepackage{tabularx}
\usepackage{hyperref}
\usepackage{enumitem}
\usepackage[acronym, nopostdot]{glossaries}

\newcolumntype{C}[1]{>{\centering\arraybackslash}p{#1}}

\newcolumntype{M}[1]{>{\centering\arraybackslash}m{#1}}
\newcolumntype{N}{@{}m{0pt}@{}}
\usepackage{fancyhdr}
\fancypagestyle{plain}{
	\fancyhf{}

	\fancyfoot[C]{\bfseries\thepage} 
}
\fancypagestyle{mainmatter}{
	\fancyhf{}

	\fancyhead[LO]{\bfseries\nouppercase\rightmark}
	\fancyhead[RE]{\bfseries\nouppercase\leftmark}
	\fancyfoot[C]{\bfseries\thepage} 
}
\fancypagestyle{frontmatter}{
	\fancyhf{}

	\fancyfoot[C]{\bfseries\thepage} 
}
\title{\bf Improving Parameter Identifiability in Household-Transmission Models Using-Genomic Data}

\usepackage{cleveref}
\makeglossaries
\definecolor{lightpink}{rgb}{1.0, 0.71, 0.76}
\usepackage{fontawesome5}

\usepackage[parfill]{parskip}

\usepackage{authblk}

\usepackage[
    backend=biber,
    style=authoryear,
    maxcitenames=1,
    mincitenames=1,
    maxbibnames=1,
    minbibnames=1,
    uniquelist=false,
    dashed=false
]{biblatex}

\title{\bf A framework for combined epidemiological-genomic inference to improve estimation of household model parameters}

\author[1]{Golsa Sayyar}
\author[2]{Joe Hilton}
\author[1]{Thomas House}

\affil[1]{Department of Mathematics, The University of Manchester, Manchester M13 9PL, UK}
\affil[2]{Manchester Centre for Health Economics, The University of Manchester, Manchester M13 9PL, UK}

\date{}

\begin{document}
	
\maketitle

\begin{abstract}
\noindent{}Models incorporating household structure, with different rates of
transmission within and between households, are widely used in infectious disease
epidemiology. These models can be calibrated using final-size data in which
transmission ordering is ignored because it does not affect the distribution
of final outbreak sizes. In particular, many distinct
transmission histories produce identical final epidemiological outcomes,
making it difficult to distinguish internal (within-household) from external
(between-household) transmission and limiting parameter identifiability.
Here, we develop a continuous-time Markov chain formulation for household
transmission dynamics in which the model state
space is expanded to include transmission graphs describing infection
direction and order, with idealised pathogen genomic data used to identify
the transmission histories compatible with observations.
We conduct simulation studies which show that incorporating genetic information
substantially concentrates the regions of high likelihood compared with models based
on epidemiological data alone. In particular, genomic data reduces the
dependence between internal and external transmission parameters,
removing the characteristic ridge associated with their weak identifiability.
These results demonstrate that graph-resolved household models enable
improved transmission inference while maintaining analytical and
computational tractability.
\end{abstract}

\section{Introduction}
	
Infectious diseases continue to pose a major threat to global health, with mathematical models playing an important role in informing public health responses \parencite{ashcroft_effectiveness_2025}. The usefulness of models, however depends critically on the accurate estimation of key epidemiological parameters such as transmission rates \parencite{Alahmadi:2020}.
	
Households are widely recognised as an important setting for transmission due to repeated close contact between individuals. As a result, household epidemic models have been used extensively to study transmission dynamics and to estimate parameters relevant to control strategies \parencite{ball1997epidemics, House:2008, ball2011household, andersson2012stochastic}.
These models commonly distinguish between transmission from outside the household and transmission between household members, and many classical results describe the distribution of final outbreak sizes under such models \parencite{ball1997epidemics}.

However, while the construction of final size distributions given a set of epidemic parameters is well-understood, the opposite process of inferring parameters from final size data is substantially more difficult. Because distinct transmission histories can generate the same observed household outcomes, household final size data alone is typically insufficient to identify the underlying transmission history, and standard household models effectively average over all compatible transmission pathways \parencite{fraser2007estimating, ball2002general}.  In the context of inference, this inability to identify a specific sequence of infection events can add substantial uncertainty to estimates of the underlying parameters. In particular, changes to transmission intensity at the within-household and between-household levels can have identical effects on final size distribution, meaning these two parameters can not be inferred independently

One natural way to address this limitation is to consider designs and analysis strategies that involve repeated observation of the household. These typically require transmission rates to vary over time, reflecting changes in epidemic pressure in the wider population \parencite{pellis2011epidemic}. While such time-inhomogeneous models are conceptually appealing and can, in principle, improve identifiability, they can lack the analytical tractability that makes homogeneous models attractive. In general, making transmission rates or other parameters explicit functions of time prevents the use of analytical methods such as matrix exponentiation, and solutions must be obtained via numerical integration or approximation techniques (\textcite{chowell2016mathematical, dureau2013capturing, martcheva2009non} and see also Appendix \ref{app2}).

In this study, we take a different but complementary approach; rather than increasing model complexity through time-dependent transmission rates, we instead incorporate an additional source of information: pathogen genomic data. 	A viral genome consists of a sequence of nucleotides. These nucleotides are represented by four letters: adenine (A), guanine (G), cytosine (C), and thymine (T) in DNA viruses, or uracil (U) instead of thymine in RNA viruses. A genetic sequence is therefore simply a string of these letters. Over the course of transmission, viral genomes accumulate mutations, meaning that the sequences observed in different individuals may differ slightly from one another. These sequences collected from infected individuals carry information about transmission relationships, with genetically similar viruses pointing to common transmission histories  \parencite{jombart2014bayesian,didelot2014bayesian, volz2013inferring, didelot2017genomic}. Genomic data have been widely used to reconstruct transmission trees and study fine-scale epidemic dynamics \parencite{treefootmouth,treereview, lau2015systematic, treeworby4014}, and have been shown to be particularly informative in household studies, where they have been used to distinguish within-household transmission from distinct external introductions, and estimate households secondary infection rate \parencite{genetichousehold}.

Here we seek to expand the set of methodological tools available in combined genomic-epidemiological inference. We propose a framework in which genomic information is used to distinguish between competing transmission graphs within a standard household epidemic model. This allows us to preserve the analytical and computational tractability of classical household models, while overcoming one of their fundamental limitations: the inability to distinguish between different transmission histories that lead to the same final outcome. As we demonstrate, even in the simplest setting of a three-person household, incorporating genetic information fundamentally alters the likelihood structure and leads to more informative inference on transmission parameters. More broadly, this approach provides a principled way to integrate pathogen genomic data into structured epidemic models without requiring fully time-dependent formulations.

In section \ref{method} (\nameref{method}), we first describe how pathogen genetic information is incorporated into our household transmission model and outline the underlying model assumptions. We then introduce graph-resolved household states, which capture transmission histories given the genomic data. Based on these states, we formulate the corresponding Kolmogorov equations and derive the likelihood function for parameter inference. We compare three scenarios: (i) a combined epidemiological and genetic model, (ii) an epidemiological-only model, and (iii) a genetic-only model. We derive the posterior distributions of the model parameters using MCMC methods for when recovery rate is fixed. Subsequently, we extend the analysis to infer all three parameters simultaneously: the recovery rate, the internal transmission rate, and the external transmission rate. Finally, we describe the procedure used to generate synthetic data from the graph-resolved household model.

We demonstrate in Section \ref{results} (\nameref{results}) that under the combined epidemiological and genetic framework, transmission parameters are more readily identifiable when information on transmission order is available and can be estimated independently. In contrast, inference based solely on epidemiological data exhibits a strong dependency between transmission parameters (Figure \ref{fig:likelihood}). Furthermore, the posterior distributions obtained from the combined model are more concentrated around the true parameter values than those obtained from the epidemiological-only model (Figure \ref{fig:posterior}). Figure \ref{genepivsgen} highlights a complementary limitation of genetic data: although genomic sequences provide information about transmission histories, they do not contain information about the duration of infectiousness. Consequently, the recovery rate remains weakly identifiable when only genetic data are available. Reliable estimation of the recovery rate therefore requires epidemiological information that determines the temporal scale of the transmission process.

Together, these findings demonstrate that epidemiological and genetic data provide distinct but complementary sources of information for transmission inference as we discuss in Section \ref{discuss} (\nameref{discuss}).

\section{Methods}\label{method}

\subsection{Setup}

In this study we seek to infer transmission parameters based on data gathered at the household level. This data consists of test results indicating whether an individual is either currently infectious or has antibodies indicative of a past infection, as well as genomic data which can be used to measure single nucleotide polymorphisms (SNP) distances between pairs of individuals. These SNP distances summarise the degree of relatedness of two individual's viral populations. This in turn can be used to quantify the likelihood of a direct epidemiological link between two cases. If two individuals in the same household are infected with viruses whose sequences differ substantially (for example, by differences at several nucleotide positions), then it is unlikely that one infected the other. Instead, it is more plausible that both individuals were infected independently from outside the household. In contrast, if the two sequences differ by only a small number of  SNPs, then transmission between the two household members is more likely.
	
\begin{figure}[t]
    \centering
    
\begin{minipage}[t]{0.48\textwidth}
        \centering
\includegraphics[width=\textwidth]{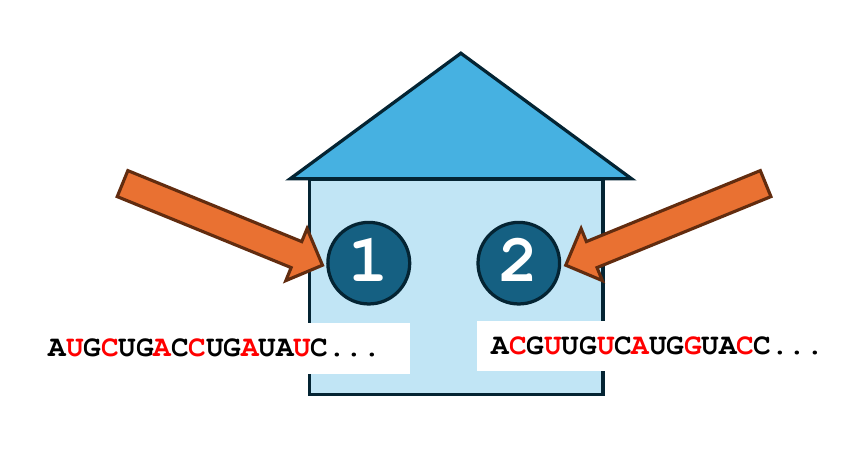}
 \subcaption*{(a)}
\end{minipage}
\begin{minipage}[t]{0.48\textwidth}
        \centering
\includegraphics[width=\textwidth]{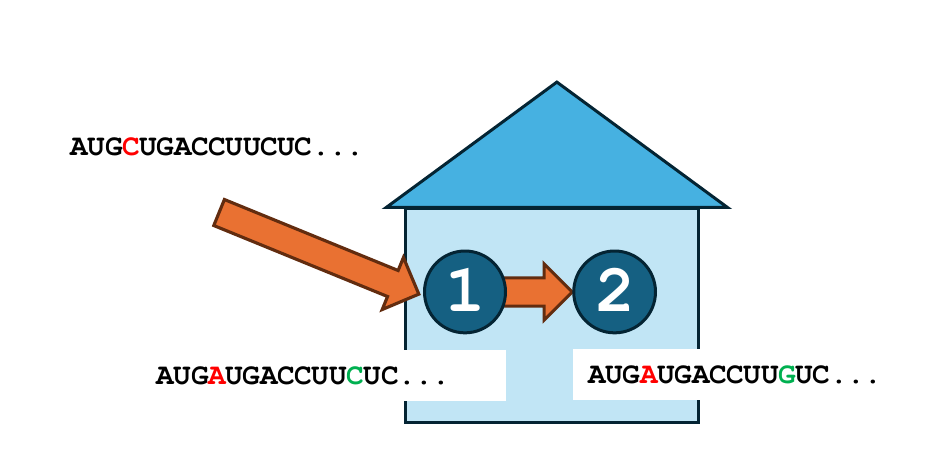}
\subcaption*{(b)}
    \end{minipage}
    \hfill
\begin{minipage}[t]{0.48\textwidth}
        \centering
\includegraphics[width=\textwidth]{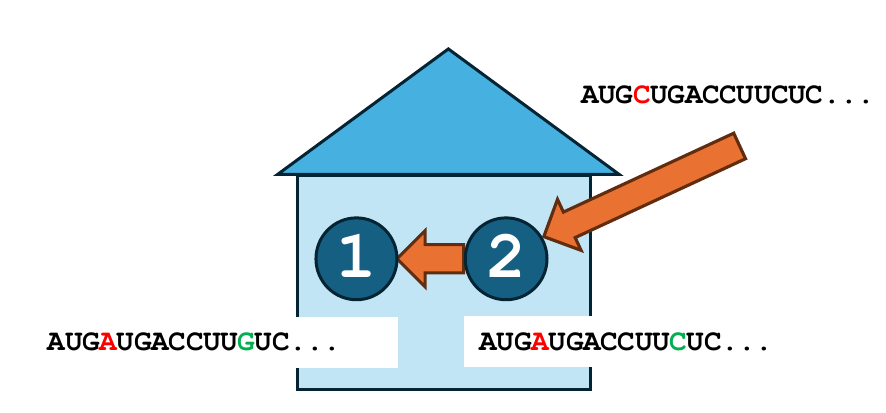}
        \subcaption*{(c)}
    \end{minipage}
    	
	\captionsetup{list=false}
\caption{Transmission scenarios in a household of size two. (a) No transmission between household members, suggesting both infections were acquired independently from outside the household. (b) and (c) The sequences are closely related, indicating possible transmission within a household. These graphs share the same final size but differ in transmission direction and infection order.}

	\label{figdirection}
\end{figure}

	Figure \ref{figdirection} illustrates this idea for a household of size two in which both individuals are infected. In panel (a), the sequences obtained from individuals A and B are very different, suggesting that there was no transmission within the household and that both infections were acquired independently from outside the household. In panels (b) and (c), where there is a single import of infection into the household followed by an onward transmission event, the two sequences are closely related, reflecting an increased likelihood that one individual infected the other within the household.

Our analysis relies on the key assumptions below about the data-gathering context for our joint epidemiological-genomic data: 
\begin{enumerate}
    \item the data are gathered in the context of a large epidemic where there is high-confidence estimates of community prevalence;
    \item the number of individuals who are directly sampled is small in relation to the ambient population;

\end{enumerate}
The first assumption ensures that, for a given household, we can define a time-stratified force of infection proportional to the community prevalence at time $t$, $\langle I(t) \rangle$; that is, the probability of an individual becoming infectious due to contact with an individual from outside their own household is proportional to $\langle I(t) \rangle$. The second assumption ensures that the probability of direct contact between any given pair of households in the sample is low, so that their infectious disease dynamics can be considered to be approximately independent of one another. 

As a simplifying assumption, we will assume that the process which generates this data is Markovian. One of the benefits of the augmented household transmission model we introduce here is that it offers a convenient way to incorporate history dependence in a Markovian setting, with the transmission tree summarising key aspects of the household-level natural history. In general, the rates of infection will be time-inhomogeneous, with the per-susceptible external infection rate given by
\begin{equation}\label{eqn:external-rate}
    \beta(t) \langle I(t) \rangle,
\end{equation}
where $\beta(t)$ is a time-dependent rate constant and the $\langle I(t) \rangle$ is the number of infections in the community. The internal infection rate between each susceptible-infectious pair is given by 
$\tau(t)$,
where $\tau(t)$ is a time-dependent transmission rate constant for members of the same household. In what follows, we will make the simplifying assumption that these rates are constant over the time interval for the study, $\mathcal{T} \subseteq [0,\infty)$, so that
\begin{equation} \label{eq:lallag}
    \beta(t) \langle I(t) \rangle =: \lambda_G, \qquad 
    \tau(t) =: \lambda_L, \qquad t \in \mathcal{T}.
\end{equation}
with subscripts $G$ and $L$ denoting ``global'' and ``local'' transmission, respectively. This assumption is mainly for simplicity, but the approach we take can be straightforwardly used for more general time-varying rates. We will consider the scenario in which the two transmission parameters $\lambda_G$ and $\lambda_L$ are unknown and are targets for inference alongside the rate of recovery, $\gamma$. In Appendix \ref{app1}, we discuss the reasons why standard final size results for household epidemic models such as in \textcite{ball1997epidemics} cannot be applied straightforwardly to this problem.

\subsection{Graph-resolved household model}\label{sec4.}

We will now define our graph-resolved continuous-time Markov chain formulation. This augments a classical continuous-time Markov chain model of infection with a graph formulation of the transmission history. Because of the dependence of SNP distance on this transmission graph, as illustrated in Figure~\ref{figdirection}, we can use this graph formulation to assign likelihood to SNP distance data, given an appropriate distribution for SNP distance conditional on the presence/absence of a shared transmission link between two cases.

We consider a closed household of fixed size $n$, indexed by individuals $1,2,\dots,n$. Each individual occupies one of three epidemiological states: Susceptible ($S$), Infectious ($I$), or Recovered ($R$). Let
$	
	X(t) = \bigl(X_1(t),\dots,X_n(t)\bigr) \in \{S,I,R\}^n
$
denote the vector of epidemiological states at time $t\ge0$, where $X_i(t)$ records the state of individual $i$. Note that unlike typical compartmental models which specify only the number or proportion of individuals in each epidemiological compartment, this formulation directly specifies the epidemiological status of each household member. The assumption of an SIR structure is made for purposes of clarity, but our basic approach can be generalised to many other compartmental structures, including structures which use Erlang or phase-type distributions to approximate non-Markovian waiting times with repeated compartments.

    In addition to the epidemiological configuration, we also track the infection history within the household by coupling this configuration with a transmission graph specifying who infected whom. We represent the transmission history as a directed graph
$
G=(V,E),
$
where the vertex set is
$
V=\{1,2,\ldots,n\},
$
and the directed edge set satisfies
$
E \subseteq ({0}\cup V)\times V.
$
Here, the special node $0$ represents infection originating from outside the household. Thus, an edge $(0,i)\in E$ indicates that individual $i$ was infected from outside the household, while an edge $(j,i)\in E$ indicates that individual $j$ infected individual $i$ within the household.

Importantly, the external source $0$ does not correspond to any specific individual and therefore carries no epidemiological state. Consequently, the household Markov process evolves only over the states of individuals indexed by $V$, while $0$ appears solely as a label for infection origin in the transmission graph.

The case of an empty edge set, $E=\varnothing$, corresponds to the absence of recorded transmission events, i.e. a purely state-based epidemiological description without inference of infection pathways.

In Figure \ref{figdirection}, for a household of size two in which both individuals are infected, three distinct transmission graphs are possible: (a) both infections are imported independently; (b) individual 1 is infected externally and subsequently infects individual 2; or (c) individual 2 is infected externally and subsequently infects individual 1. These graphs share the same final size but differ in transmission direction and infection order.

Let $\mathcal G$ denote the set of all  transmission
graphs describing infection histories within the household . The household epidemic is described jointly by the
epidemiological configuration and the transmission graph.
We therefore define the graph--resolved household state as
$
Z(t) = (X_1(t),\ldots{},X_n(t),G(t)),
$
where we will index state labels with the individual number for later notational transparency, so that each
$X_i(t)\in\{S_i,I_i,R_i\}$, with $\{S,I,R\}^n = \{S_1,I_1,R_1\} \times \cdots \times \{S_n,I_n,R_n\}$,
and $G(t)\in\mathcal G$. 
The process then evolves on the finite state space
$
\mathcal Z
=
\{S,I,R\}^n \times \mathcal G ,
$  
with events and rates
\begin{align}
(X_1, \ldots, S_i, \ldots, X_n, G) & \rightarrow (X_1, \ldots, I_i, \ldots, X_n, G\cup (0,i))
	& & \text{at rate } \lambda_G \text{ ,}\nonumber \\
(X_1, \ldots, S_i, \ldots, X_n, G) & \rightarrow (X_1, \ldots, I_i, \ldots, X_n, G\cup (j,i))
	& & \text{at rate } \lambda_L \mathbb{I}_{\{X_j = I_j\}} \text{ ,} \nonumber \\
(X_1, \ldots, I_i, \ldots, X_n, G) & \rightarrow (X_1, \ldots, R_i, \ldots, X_n, G)
	& & \text{at rate } \gamma \text{ ,} \label{eq:GXmodel}
\end{align}
for each $i,j \in V$.

We partition the graph space according to the cumulative number of within-household infection events. Define
\[
\mathcal G
=
\bigcup_{k=-1}^{n-1} \mathcal G_k,
\]
where \(\mathcal{G}_{-1}\) consists of the transmission graph with no infection events, i.e.
\[
\mathcal{G}_{-1}
=
\left\{ G : E(G) = \varnothing \right\}.
\]

For $k>0$, $\mathcal {G}_k$ denotes the set of transmission graphs
involving $k$ infection events within
the household. As the within-household transmission dynamics evolve, a household's graph may transition both within and between these sets. The class $\mathcal {G}_0$ consists of graphs in which all
observed infections arise from external introduction,
that is, for $G \in \mathcal {G}_0$, 
$
E(G)\subseteq
\{(0,i): i\in V\}.
$
Hence every infection edge for graphs in $\mathcal{G}_0$ originates from the external node $0$.  Graphs in $\mathcal G_1$ are obtained from graphs in
$\mathcal G_0$ by adding exactly one internal transmission
edge,
\[
G^{(1)} = G \cup (j,i),
\qquad
G\in\mathcal G_0 , \quad 
i,j\in V ,
\]
where individual $j$ has previously been infected externally and individual $i$ had not previously been infected. The resulting graph satisfies
$
G^{(1)} \in \mathcal G_1 .
$ Adding another external infection would reproduce a graph
already contained in $\mathcal G_1$; therefore the transition
from $\mathcal G_0$ to $\mathcal G_1$ necessarily corresponds to the first within-household transmission event. More generally, graphs in $\mathcal G_{k\ge1}$
contain at least $k+1$ total infection events and are generated
recursively through admissible infection extensions:
\[
G^{(k+1)} = G \cup e,
\qquad
G\in\mathcal G_{k},
\]
where the added edge $e$ may be
$
e\in
\{(0,i)\}
$, due to external infection,
or
$
e\in
\{(j,i): j\in V\}
$, as a result of internal infection,
provided the infector is already infected in $G$ and the infectee is not. Then
$
G^{(k+1)} \in \mathcal G_{k}
$ or $
G^{(k+1)} \in \mathcal G_{k+1},
$ respectively.

This decomposition of the graph space induces a
corresponding partition of the joint state space.
We define
\[
\mathcal Z_k
=
\{(X,G)\in\mathcal Z : G\in \mathcal G_k \},
\qquad
\mathcal Z
=
\bigcup_{k=-1}^{n-1}\mathcal Z_k, 
\]
where $\mathcal Z_{-1}$ corresponds to the state with no infection transmission, i.e., in susceptible state, where all individuals are susceptible.

\subsection{Transmission-graph household model}
	
The graph-augmented household epidemic is modelled as a continuous-time Markov process. Note that the state-space $\mathcal{Z}$ has finite cardinality, which we call $N = |\mathcal{Z}|$.  Let $P(t)$ denote the length-$N$ vector of probabilities over all possible graph-labelled household states at time $t$. Note that these states are not defined only by the epidemiological configuration (e.g.\ $(S,I)$, $(I,R)$, $(R,R)$ \ldots{} for $n=2$), but by an array consisting of epidemiological state and transmission graph. As a result, two households that appear epidemiologically identical but arise from different transmission histories are treated as distinct states in the model.
	
The time evolution of $P(t)$ is governed by the Chapman-Kolmogorov equations
	\begin{equation} \label{eq:CK}
		\frac{\mathrm{d}{P}(t)}{\mathrm{d}t} = Q(t) P(t),
	\end{equation}
	where $Q(t)$ is the transition rate matrix defined on this expanded state space. As in classical household models, $Q$ can be decomposed in general as
\begin{equation}
    Q(t) = \tau(t) Q_L + \beta(t) \langle I(t) \rangle Q_G + \gamma Q_R,
\end{equation}
where as in \eqref{eq:lallag}  we will typically make a simplifying assumption of time homogeneity so that $\beta(t) \langle I(t) \rangle = \lambda_G$ and $\tau(t) = \lambda_L$ during the time period of interest. Note that in contrast to previous models based on ODEs of the form \eqref{eq:CK} as in e.g.\ \textcite{kinyanjui2018scabies}, the entries of $Q$ are constructed so as to preserve information about the direction and order of transmission. In particular, transitions in different directions but involving the same individuals (e.g.\ $i \rightarrow j$ versus $j \rightarrow i$) are represented as distinct transitions. 

Precisely, the elements of $Q$ are obtained using standard methods for continuous-time Markov chains \parencite{norris97} from the transitions and rates of the graph-labelled process given in
\eqref{eq:GXmodel}. Since this process has a finite number of states, say $N$, this means we construct a map $\phi : \mathcal{Z} \rightarrow \{1,\ldots,N\}$. Loosely speaking, where there is a transition from state $z$ to state $z'$ at rate $r$, we subtract $r$ from $Q_{\phi(z), \phi(z)}$ and add $r$ to $Q_{\phi(z'), \phi(z)}$. We will now discuss the different events and rates for the transmission-graph household model.

If individual $i$ is susceptible, it can be infected from outside the household 
\[
(X_1, \ldots, S_i, \ldots, X_n, G)
\longrightarrow
(X_1, \ldots, I_i, \ldots, X_n, G\cup (0,i))
\]
occurs at rate $\lambda_G$.
Since this event introduces one additional infection edge
into the transmission graph, it increases the cumulative
number of infection events by one. Hence the transition is
$
\mathcal Z_k \to \mathcal Z_{k}
$ for $k \in \{0,1,..,n-1\}$ and for $k={-1}$ it is
$
\mathcal Z_{-1} \to \mathcal Z_{0}
$.

If $j$ is infectious and $i$ susceptible,
\[
(X_1, \ldots, S_i, \ldots, X_n, G)
\longrightarrow
(X_1, \ldots, I_i, \ldots, X_n, G\cup (j,i))
\]
occurs at rate $\lambda_L$. This event likewise adds a new infection edge to the graph, so the
transition is $\mathcal Z_k \to \mathcal Z_{k+1}$ for $k \in {0,1,..,n-1}$.

Recovery transitions modify only the epidemiological state:
\[
(X_1, \ldots, I_i, \ldots, X_n, G)
\longrightarrow
(X_1, \ldots, R_i, \ldots, X_n, G)
\]
at rate $\gamma$, and therefore the transition is
$
\mathcal Z_k \to \mathcal Z_k .
$

Since the matrix $Q $ is time-homogeneous, the system admits the formal solution as a matrix exponential as in Appendix \ref{app2}, Equation \eqref{eq2}. We would like our methods to be applicable to cases in which the external force of infection is time-varying and have therefore implemented code that solves the Chapman–Kolmogorov equations using stepping methods, as motivated in Appendix \ref{app2}.

\subsection{Likelihood formulation}

 We will assume that each household is observed at a finite set of observation times in a vector
$T = (t_1, t_2, \dots, t_m),$
where $t_1 < t_2 < \cdots < t_m$.  For each household $h$, the full trajectory of states evaluated at these observation times is $
Z^{(h)}(T)=\left(Z^{(h)}(t_1),\dots,Z^{(h)}(t_m)\right)
$
where each $Z^{(h)}(t_i) \in \mathcal{Z}$, and the epidemiological component may be written as the matrix
$
X^{(h)}
\in \{S,I,R\}^{n \times m},
$
whose \((i,k)\)-th entry records the state of individual \(i\) in household \(h\)
at time \(t_k\). For fixed parameters \(\theta\), the Chapman-Kolmogorov system \eqref{eq:CK} can be used to calculate terms of the form
$
\Pr(Z \mid \theta),
$
the probability of each graph-labelled trajectory given $\theta$. We note that such calculations will involve some resetting of the initial conditions at each observation point. Dealing with this is in general quite complicated. Here, however, we will simplify analysis by only considering simulated data for which $m=2$ (but different observation times, one with $t_{m}=2$ and one with $t_{m}=10$) meaning we do not need to reset initial equations.

We will, however, build on existing work by considering the case in which the full state \(Z^{(h)}\) is not directly observed, and instead, we observe a function of the full state. Let $
\Phi : \mathcal{Z} \to \mathcal{Y}
$ be an observation map into an observation space \(\mathcal{Y}\). The observed data for household \(h\) are therefore
$
Y^{(h)} = \Phi(Z^{(h)})$, where the function $\Phi$ acts element-wise on its argument, and the likelihood of parameters $\theta$ given $Y^{(h)}$ is then
\[
\Pr\!\left(Y^{(h)} \mid \theta\right)
=
\sum_{\substack{Z \in \mathcal{Z} \\ \mathcal{Z}: \Phi(Z)=Y^{(h)}}}
\Pr(Z \mid \theta).
\]
i.e., the sum of the probabilities of graph-labelled states that yield the observed data under the observation function $\Phi$.

For a set of $N_{\text{HH}}$ households which undergo independent transmission dynamics, the full likelihood is given by

\begin{equation}\label{eqn:likelihood-sum}
\mathcal{L}(\theta)
=
\prod_{h=1}^{N_{\text{HH}}}
\Pr(Y^{(h)} \mid \theta).
\end{equation}

To assess the respective contributions of genomic and epidemiological data to household parameter inference, we will carry out likelihood calculations based on three different choices of the
observation map \(\Phi\), each of which conserves different aspects of the combined genomic-epidemiological state space.

\textbf{(i) Full transmission information (epidemiological + genetic data).}

In the primary analysis, we assume that the full graph-labelled state
is observed. The observation map is therefore the identity:
$
\Phi_{\text{full}}(Z) = Z,
$
and so
$
Y^{(h)} = Z^{(h)}.
$
Because the true underlying state uniquely determines the observation in this case, the sum in Equation~\ref{eqn:likelihood-sum} has only a single term and thus

\[
\Pr(Y^{(h)} \mid \theta)
=
\Pr(Z^{(h)} \mid \theta).
\]

The joint likelihood function is then

\[
\mathcal{L}_{\text{full}}(\theta)
=
\prod_{h=1}^{N_{\text{HH}}}
\Pr(Z^{(h)} \mid \theta).
\]

\textbf{(ii) Genetic data only.}
If only genetic information is available, infectious
and recovered individuals can not be distinguished, although susceptible individuals can be identified by their absence from the set of transmission tree vertices. States that differ only by replacing \(I\) with \(R\) are therefore equivalent.

Let \(\tilde{X}\) denote the configuration obtained from \(X\)
by collapsing the classes \(I\) and \(R\) into a single category.
The observation map is then
$
\Phi_{\text{gen}}(Z) = \tilde{X}.
$
Thus,
$
Y^{(h)} = \tilde{X}^{(h)}.
$
The likelihood contribution becomes
\[
\Pr(Y^{(h)} \mid \theta)
=
\sum_{Z \in \mathcal{Z} : \Phi_{\text{gen}}(Z)=Y^{(h)}}
\Pr(Z \mid \theta).
\]
Hence,
\[
\mathcal{L}_{\text{gen}}(\theta)
=
\prod_{h=1}^{N_{\text{HH}}}
\sum_{Z : \Phi_{\text{gen}}(Z)=Y^{(h)}}
\Pr(Z \mid \theta),
\]
which corresponds to summing graph-resolved probabilities over
all states that are identical up to the \(I/R\) distinction.

\textbf{(iii) Epidemiological data only.}
If only the epidemiological configuration is observed, then the observation map projects onto the epidemiological component:
$
\Phi_{\text{epi}}(Z) = X,
\quad
\text{for } Z=(X,G).
$
Thus,
$
Y^{(h)} = X^{(h)}.
$
Since multiple transmission graphs may produce the same epidemiological states, the likelihood contribution is obtained by summing over all compatible graph-labelled states:

\[
\Pr(Y^{(h)} \mid \theta)
=
\sum_{\substack{Z \in \mathcal{Z} \\ \mathcal{Z}: \Phi(Z)=X^{(h)}}}
\Pr(Z \mid \theta),
\]

where \(X(Z)\) denotes the epidemiological component of \(Z\).
The corresponding likelihood is

\[
\mathcal{L}_{\text{epi}}(\theta)
=
\prod_{h=1}^{N_{\text{HH}}}
\sum_{Z : \, \Phi_{epi}(Z)=X^{(h)}}
\Pr(Z \mid \theta).
\]

In practice, this corresponds to aggregating graph-resolved
probabilities into epidemiological classes.

\subsection{Posterior distributions}

To more fully assess the performance of our proposed inference method, we will consider two inference problems, which we consider in a Bayesian framework, seeking to analyse posterior distributions over parameters $\theta$ given data $Y$.

\textbf{(i) Fixed recovery rate.}
In the baseline analysis, the recovery rate is assumed known and fixed, and we infer only the transmission parameters.
The parameter vector is therefore
$
\theta = (\lambda_L,\lambda_G),
$
and the posterior distribution is

\[
\pi(\lambda_L,\lambda_G \mid Y, \gamma)
\propto
\mathcal{L}(\lambda_L,\lambda_G \mid Y, \gamma)
\,
\pi(\lambda_L,\lambda_G).
\]

\textbf{(ii) Joint inference of recovery and transmission rates.}
In the extended analysis, we infer all parameters jointly,
$
\theta = (\lambda_L,\lambda_G,\gamma),
$
with posterior

\[
\pi(\lambda_L,\lambda_G,\gamma \mid Y)
\propto
\mathcal{L}(\lambda_L,\lambda_G,\gamma \mid Y)
\,
\pi(\lambda_L,\lambda_G,\gamma).
\]

In practice, lognormal priors are often a natural choice for positive rate parameters, as they restrict the parameters to positive values while allowing substantial variation over several orders of magnitude. However, for this simple example, we use independent uniform priors on $[0,1]$ for all parameters:
$\theta_i \sim \mathrm{Uniform}(0,1)$ for $i=1,\ldots,d$,
where $d$ denotes the number of parameters; for example, $d=3$ corresponds to
$\theta=(\lambda_L,\lambda_G,\gamma)$.

The joint prior density is therefore
$\pi(\theta)=\prod_{i=1}^{d}\pi_i(\theta_i),$
where each marginal density is given by
\[
\pi_i(\theta_i)=
\begin{cases}
1, & 0\leq\theta_i\leq1,\\
0, & \text{otherwise}.
\end{cases}
\]

\subsection{Synthetic data generation}\label{sec:validation}

To validate our graph-augmented inference method, we conducted experiments in which we attempted to estimate transmission parameters from synthetic data generated with known parameters $\theta$. We will denote synthetic datasets consisting of multiple household samples $Y^{(h)}$ by $\Upsilon$.

Synthetic datasets were generated from the proposed graph-resolved
household model under fixed true parameter values
$\lambda_G=\lambda_L=0.5$.
To assess the impact of the relative timings of testing and infectious periods on our method, we considered scenarios in which the true recovery
rate was set to $\gamma=0.5$ (infectious period 2), and to $\gamma=0.1$  (infectious period 10).  These parameter choices ensure that 
internal and external transmission occur at comparable rates
against a background of overall growing infection, meaning that
the two routes of transmission can not be immediately distinguished by timing. Households were all assumed fully susceptible at an initial time $t_1$ then observed at a fixed final observation time $t_{m=2}$ chosen
so that all three epidemiological states ($S$, $I$, and $R$) were present with
non-negligible probability under the true parameter values.
In particular, our assumption of a fixed constant import rate $\lambda_G$ means that late observation times would result in most households reaching the absorbing state $(R, R)$, thereby reducing the information available for parameter inference (see Figure \ref{fig:likelihood2}).

To generate a synthetic household dataset given parameters $\theta^*$, the Kolmogorov forward equations were
solved to obtain the probability distribution over all admissible
graph-labelled household outcomes,
$Pr(z \mid \theta^\ast)$ for $z \in \mathcal Z$,
where $\mathcal Z=\{z_1,\dots,z_N\}$ is the finite state space of graph-resolved household configuration. The size of this state space depends on household size, for instance, for households of size $n=2$ we obtain $N=17$ admissible graph-labelled states, while for households of size $n=3$ the expanded state space contains $N=171$ states.

We can then generate a synthetic dataset by drawing independent samples from the model-derived probability distribution
$Pr(\cdot \mid \theta^\ast)$, and then applying the appropriate observation function $\Phi$ to each sample.

\section{Results}\label{results}
    
\subsection{Likelihood and posterior behaviour under different data scenarios}\label{Sec. 5}

To estimate the transmission parameters $\theta$ from a synthetic dataset $\Upsilon$, we performed $10^4$ MCMC iterations, discarding the first
$10^3$ as burn-in. Each simulation consisted of $N_{\mathrm{HH}} = 200$ independent
households, each of size $n=3$.
	
\begin{figure}[H]
		\centering

\includegraphics[width=0.98\textwidth]{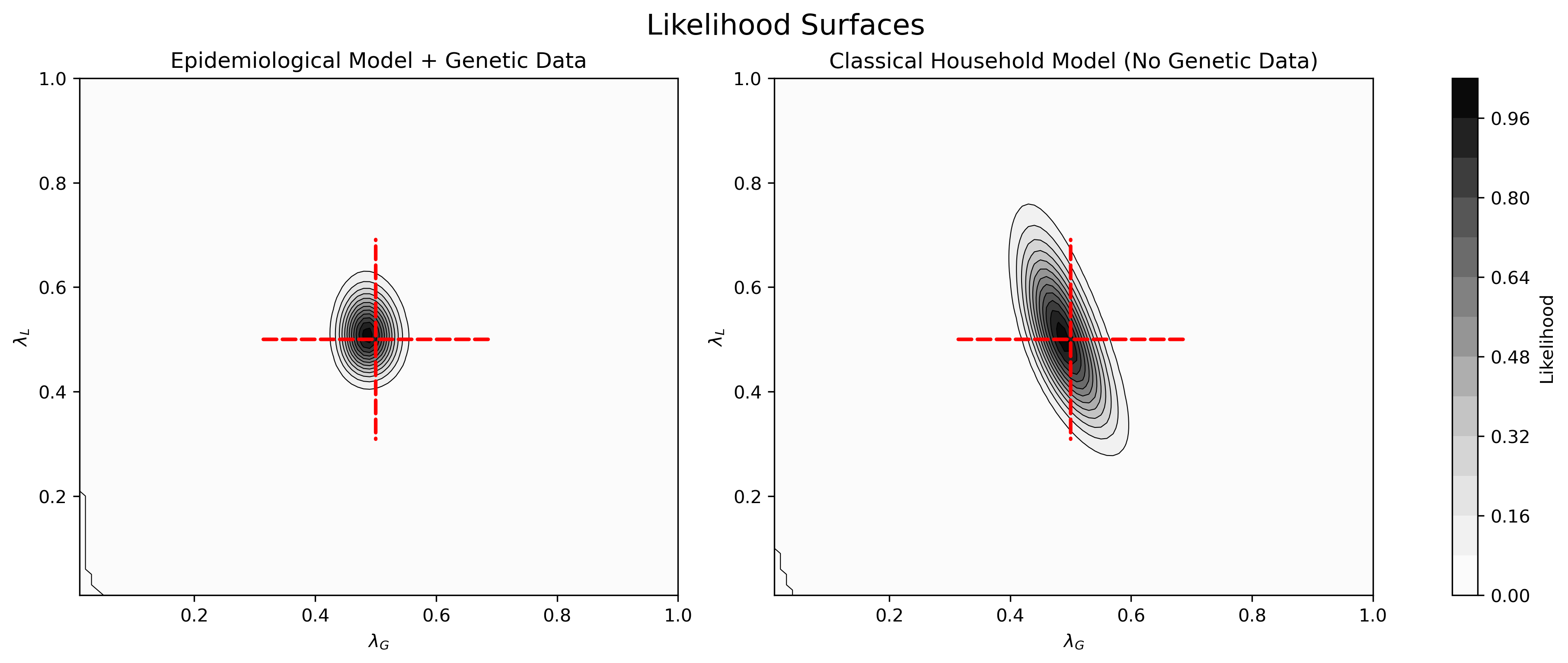}
	\caption{Likelihood surfaces for transmission parameters with (left) and without (right) genetic information. Incorporating genetic data yields a well-identified likelihood surface, whereas the model without genetic information exhibits a ridge indicating parameter non-identifiability. The dashed lines intersect at the true parameter values, $\lambda_G = \lambda_L = 0.5$, with recovery rate fixed at $\gamma = 0.5$. Results are based on simulated data from 200 households of size 3.}    
	
\label{fig:likelihood}
	\end{figure}

Figure~\ref{fig:likelihood} compares likelihood surfaces for the transmission parameters under two modelling approaches:  
	(i) a model that incorporates genetic information to distinguish transmission graphs, and  
	(ii) a model based only on epidemiological data, without genetic information,
given the same synthetic dataset $\Upsilon$.
	
When genetic information is included (left panel), the likelihood surface is sharply concentrated around the true parameter values, with an approximately circular high-likelihood region and very low correlation between the parameters. This indicates that the two parameters are well identified and can be estimated independently with high precision.
	
In settings where genetic data are not available (right panel), the likelihood surface shows a pronounced ridge structure. The high-likelihood region lies along a diagonal curve, indicating strong correlation between the parameters. This reflects the well-known identifiability problem in household models based solely on final outcomes: multiple combinations of internal and external transmission rates produce indistinguishable likelihoods. Comparable ridge-like likelihood structures
have also been reported when estimating within- and between-partnership
transmission rates in stochastic models of HIV transmission
\parencite{munoz2026inference}.
	
For Figures~\ref{fig:likelihood} and \ref{fig:posterior} the true recovery
rate was set to $\gamma=0.5$, while in
Figure~\ref{fig:likelihood2} a smaller recovery rate $\gamma=0.1$
was considered in order to examine inference under much slower recovery. 
Figure \ref{fig:likelihood2} compares likelihood surfaces obtained
at an early observation time ($t_2=2$) and a late observation time
($t_2=10$) while fixing the transmission parameters at
$\lambda_G=\lambda_L=0.5$ and choosing a relatively slow recovery rate
$\gamma=0.1$.
The smaller recovery rate increases the mean infectious period,
thereby extending the time scale of infection and recovery.
At the same time, the resulting increase in transmission potential
accelerates epidemic spread, allowing outbreaks to progress
substantially before observation.

At the earlier time point, many households remain in transient
epidemiological states containing mixtures of susceptible, infectious,
and recovered individuals.
These partially evolved outbreaks retain information about the relative
contributions of internal and external transmission, resulting in a
well-localised likelihood surface.

In contrast, when households are observed at the later time
$t_2=10$, a large proportion of infected
individuals have already progressed to the recovery state.
As a consequence, temporal information about how infection
occurred is lost. In particular, once individuals
have recovered, it becomes difficult to determine whether their
infection originated from external exposure or from
within-household transmission. 

When only epidemiological outcomes are observed, recovered
individuals are indistinguishable regardless
of infection source. Different combinations of the external
and within-household transmission rates therefore produce
similar observable household states, leading to a substantial
ridge structure observed in the likelihood surface.

Notably, identifiability depends not only on observation
time but on its relation to the epidemic time scale determined
by $\lambda_G$, $\lambda_L$, and $\gamma$.
Although in both Figure \ref{fig:likelihood} and the present setting the observation
time satisfies $t_2 \approx 1/\gamma$
($t_m=2$ when $\gamma=0.5$ and $t_2=10$ when $\gamma=0.1$), corresponding to the mean infection period,
smaller values of $\gamma$ increase the basic reproduction
number, allowing outbreaks to progress further before
observation.
This results in more recovered individuals and reduced
information about infection origin, producing a more pronounced
ridge in the likelihood surface.
 
Although this loss of information also occurs when genetic
data are available, genomic similarity between infections
provides additional constraints on plausible transmission
pathways. Therefore, genetic information partially
preserves transmission structure even at later observation
times, reducing parameter confounding and improving
identifiability. 
Consequently, the likelihood surface develops a ridge
structure, reflecting reduced parameter identifiability.

This comparison demonstrates that identifiability depends not only on
model structure or data type, but also critically on the timing of
observation relative to the epidemic time scale.

    \begin{figure}[H]
    \centering

    % Top row
    \begin{subfigure}{0.94\textwidth}
        \centering
        \includegraphics[width=\linewidth]{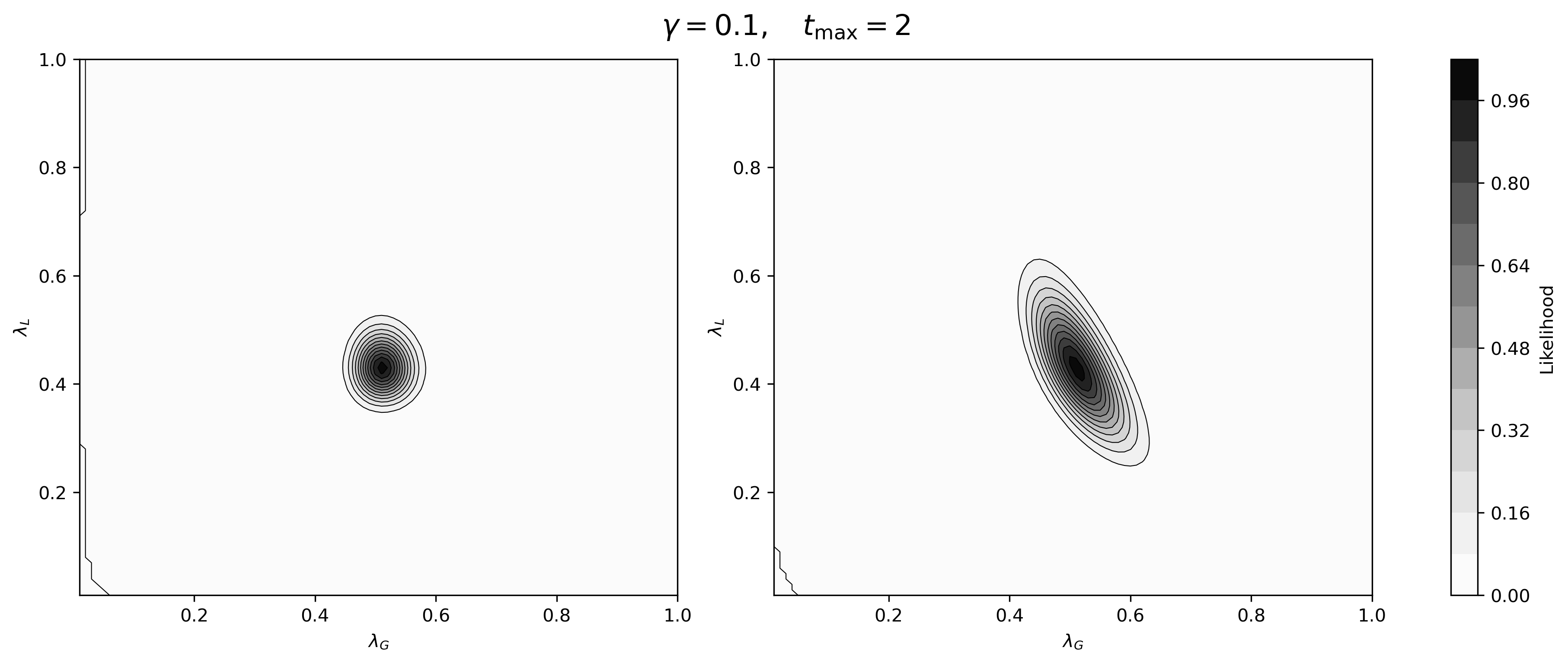}
    \end{subfigure}

    \vspace{0.5em}

    % Bottom row
    \begin{subfigure}{0.92\textwidth}
        \centering
        \includegraphics[width=\linewidth]{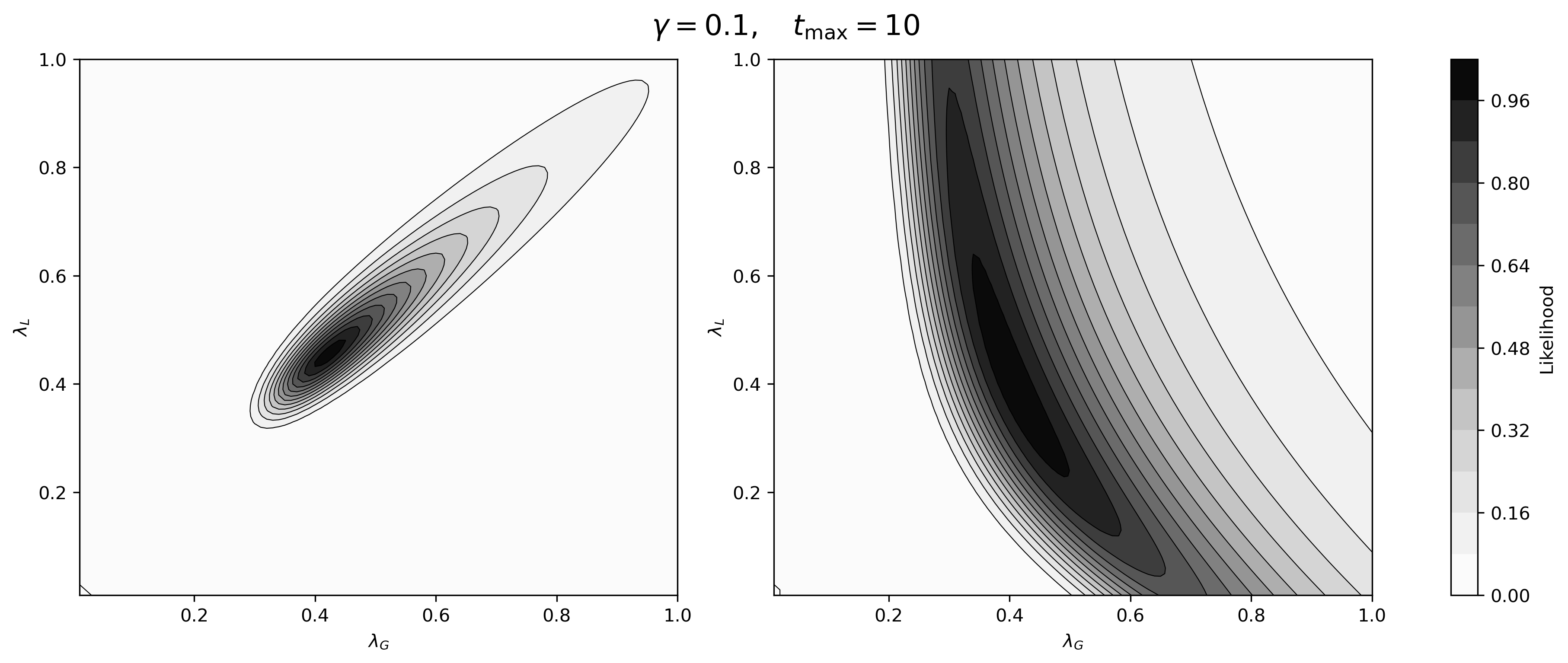}
    \end{subfigure}
    \end{figure}
\begin{figure}[H]
    \ContinuedFloat
    \caption{Likelihood surfaces for fixed transmission parameters
$\lambda_G=\lambda_L=0.5$ and recovery rate $\gamma=0.1$
evaluated at two observation times with (left) and without (right) genetic information.
Top: early observation ($t_m=2$), where transient epidemic
states preserve information about transmission mechanisms.
Bottom: late observation ($t_m=10$), where households approach
final-size configurations and parameter identifiability
deteriorates.
}
\label{fig:likelihood2}
\end{figure}

Figure~\ref{fig:posterior} shows posterior distributions obtained from the MCMC algorithm for both modelling approaches. When genetic information is incorporated (top row), the posterior distributions are tightly concentrated around the true parameter values, demonstrating accurate and precise inference.
	
In contrast, when genetic information is excluded (bottom row), the posterior distributions are substantially broader. Although the marginal posteriors still cover the true values, uncertainty is greatly increased and the joint posterior exhibits strong dependence between parameters, consistent with the ridge structure observed in the likelihood surface.
	
Together, these results demonstrate that incorporating genetic information into the household modelling framework substantially improves identifiability and enables more reliable estimation of transmission parameters, even without introducing explicit time dependence into the model.

\begin{figure}[H]
		\centering
\includegraphics[width=1\textwidth]{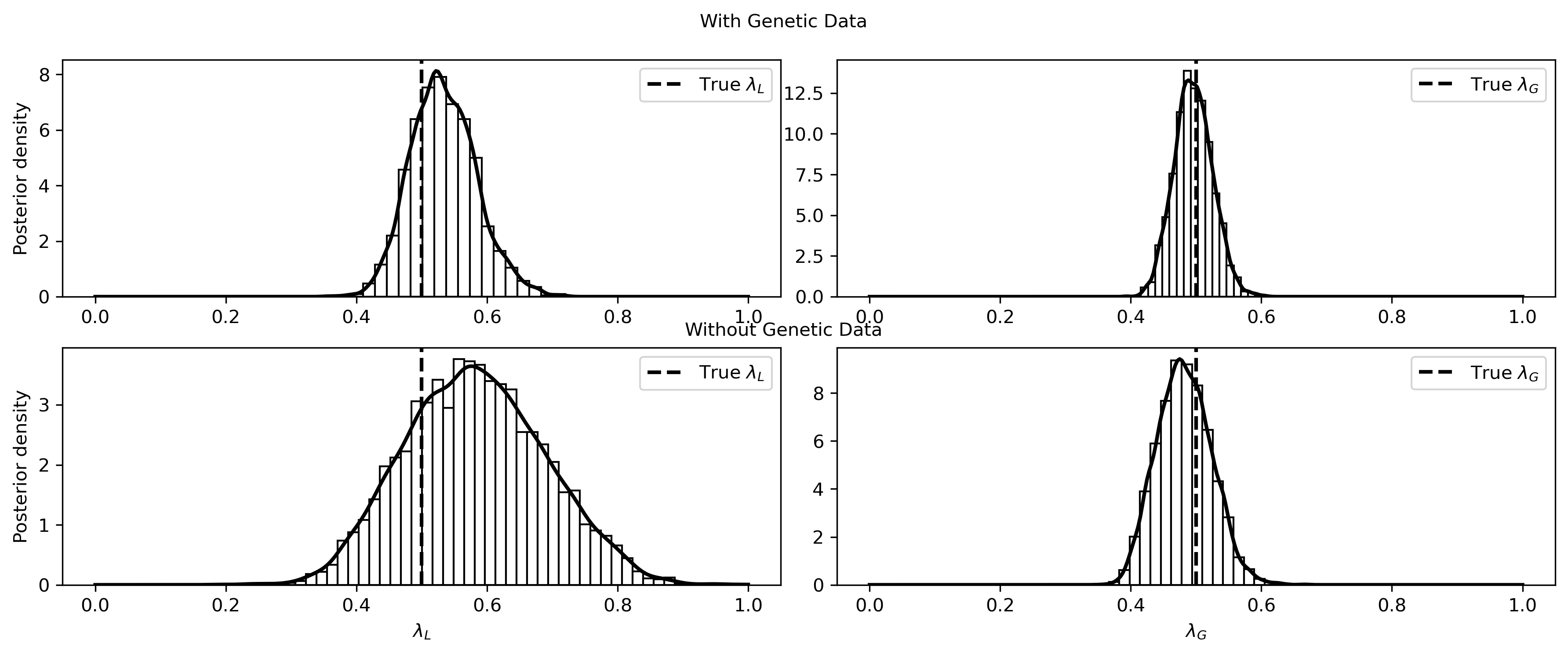}
\end{figure}

\begin{center}
\captionof{figure}{
Posterior distributions obtained from MCMC with (top row) and without (bottom row) genetic information. Incorporating genetic data leads to tighter posterior concentration and improved parameter inference. The true parameter values are $\lambda_G = \lambda_L = 0.5$ with the recovery rate fixed at $\gamma = 0.5$. Results are shown for 10,000 MCMC simulation from 200 households of size 3.}
\label{fig:posterior}
\end{center}

\subsection{Inferring the recovery rate and the role of epidemiological information}

We performed further analyses to assess two extensions of our framework. 
First, we relaxed the assumption of a fixed recovery rate and instead inferred 
$\gamma$ jointly with the transmission parameters $\lambda_G$ and $\lambda_L$. Second, we compared inference 
under three different data scenarios: 
(i) combined epidemiological and genetic information, 
(ii) genetic information alone, and (iii) epidemiological information only.

Figure \ref{genepivsgen} summarises the resulting  
likelihood surfaces. When both epidemiological outcomes and genomic data are 
available (top row), the recovery rate $\gamma$ becomes identifiable and shows only weak dependence with the transmission parameters.  In this case, epidemiological information provides the necessary time-scale, 
while genomic data help resolve transmission structure within the household.

In contrast, when only genetic information is available (middle row), the 
recovery rate is not identifiable. Although genomic sequences provide information 
about who infected whom, they do not determine the  duration of 
infectiousness, as a result, $\gamma$ remains weakly informed by the data.

Finally, when only epidemiological information is used (bottom row), 
the recovery rate can be identified provided that one of the transmission 
rates is fixed. However, as shown previously in Figure~\ref{fig:likelihood}, 
the within-household and external transmission rates remain strongly 
correlated in this setting. In particular, different values of 
$\lambda_L$ and $\lambda_G$ can generate very similar likelihood values, 
so that the transmission parameters are not separately identifiable 
even though $\gamma$ is informed by the temporal data.

Overall, these results highlight that genomic data alone are insufficient to 
identify all epidemic parameters, particularly those governing the time scale 
of infection and recovery. However, when combined with even limited 
epidemiological information, genetic data substantially improve identifiability 
of transmission parameters and enable joint inference of both transmission and 
recovery dynamics.

\begin{figure}[H]
    \centering
\includegraphics[width=0.97\textwidth]{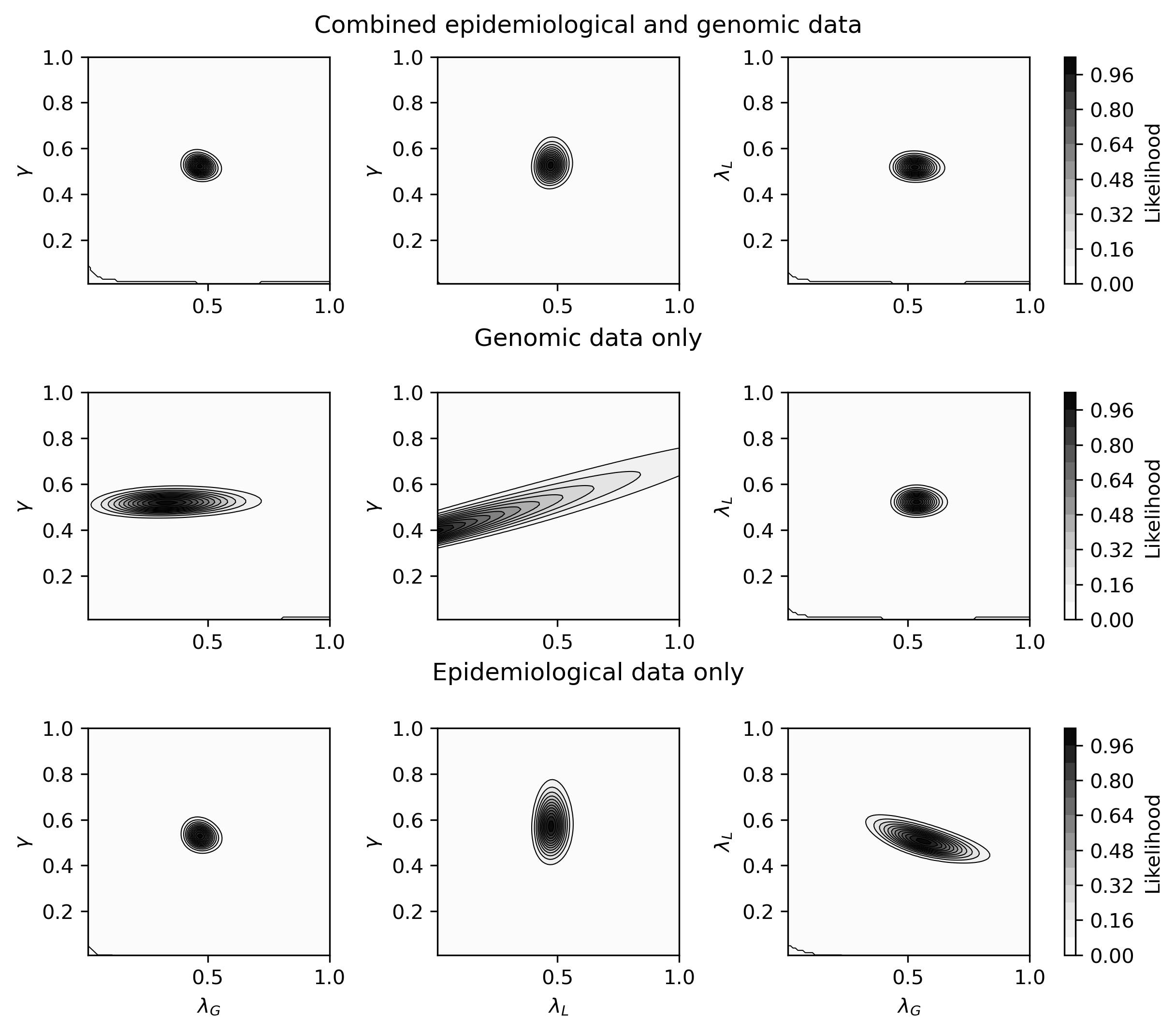}
\end{figure}

\begin{center}
\captionof{figure}{
Likelihood surfaces comparing inference under two data scenarios:
combined epidemiological and genetic information (top row) versus genetic 
information alone (bottom row). In both settings, the transmission rates 
$\lambda_L$, $\lambda_G$ and the recovery rate $\gamma$ are jointly inferred. 
The true parameter values are $\lambda_L = \lambda_G = \gamma = 0.5$.
}
\label{genepivsgen}
\end{center}

\section{Discussion} \label{discuss}

Inference for household transmission models based solely on epidemiological (e.g.\ final size) data is inherently limited by the fact that transmission pathways are unobserved. Many distinct infection histories can produce identical observations, making it impossible to identify who infected whom or to reliably separate internal from external transmission using epidemiological data alone.

In this paper, we refine the state representation of the model to include transmission routes explicitly in a way that might be identifiable from a simplification of genomic data. We preserve the standard time-homogeneous continuous-time Markov chain formulation, but expand the state space to distinguish between alternative transmission graphs that are otherwise aggregated in classical household
models. Genetic information is then used to restrict the set of plausible transmission graphs. In this way, we recover transmission-direction information without abandoning the analytical and computational advantages of homogeneous household models.

 The crucial advantage of incorporating genetic data is that it allows us to distinguish between different transmission trees that are treated as equivalent in standard household epidemic models. Classical models are typically constructed so that only the number of infections matters, not the order in which they occurred or the direction of transmission. As a result, transmission histories that lead to the same final size are averaged over and cannot be distinguished using case count data alone.
	
On the other hand, when genetic information is available, the direction of transmission can become partially identifiable. Genomic similarity may not of course always uniquely determine who infected whom, but it will generally restrict the transmission graphs that are compatible with the observed sequences. Transmission histories that would be indistinguishable under epidemiological data alone may become more or less plausible once genetic similarity is taken into account. For example, in a two-person household where one individual infects the other, genetic data can indicate which of the two was more likely to be infected first. This additional structure is scientifically meaningful: although the final number of infections in the household does not depend on who infected whom, knowledge of the transmission direction provides much richer information about the underlying epidemiological process. From a statistical perspective, this is important because it can lead to more accurate inference of key model parameters.

 \cite{kenah2016molecular} developed algorithms to enumerate transmission trees and showed that these can be used to calculate likelihoods incorporating both epidemiological data and a phylogeny. However, in their framework, the direction of transmission was not fully considered. For example, in Figure 1 of their paper, they consider only two possible transmission trees for a household size 2 when A is the source of infection: one in which A infects both B and C directly, and another one A infects B, who then infects C. By contrast, in our approach, we distinguish between the transmission sequences $A \rightarrow B \rightarrow C$ and $A \rightarrow C \rightarrow B$. Nevertheless, under their assumptions, they showed that incorporating a phylogeny into an epidemiological model produced more efficient estimates of parameters, though less efficient when combining epidemiological data with fully resolved transmission trees (Table 1-3). Our transmission-graph formulation is consistent with these findings.
In addition, we demonstrate that incorporating genetic information substantially improves identifiability of transmission parameters (Figure \ref{fig:posterior}) and leads to more informative likelihood surfaces compared with models based on epidemiological outcomes alone (Figure \ref{fig:likelihood}). Importantly, this improvement is achieved without requiring fully time-resolved models, thereby retaining mathematical and computational tractability while still capturing key features of the underlying transmission process.

A further insight is provided by Figure~\ref{fig:likelihood2},
which illustrates the influence of observation time on parameter
identifiability. When households are
observed early in the outbreak, transient epidemiological states
retain information about how infection was introduced,
leading to a well-localised likelihood surface. In contrast,
at later observation times most individuals have recovered,
making infections emerging from external and internal
transmission increasingly difficult to distinguish.
As a result, the likelihood develops a pronounced ridge structure.
From this figure and Figure \ref{fig:likelihood}, it can also be noticed that identifiability  depends not only on observation time but on
its relation to the epidemic time scale determined by
$\lambda_G$, $\lambda_L$, and $\gamma$. Although in both
figures
$t_m \approx 1/\gamma$, smaller values of $\gamma$ increase the basic
reproduction number, allowing outbreaks to progress further before
observation and thereby reducing information about infection origin.

Furthermore, as expected, Figure \ref{genepivsgen} shows that genetic data alone 
are not sufficient to reliably estimate the recovery rate. While genomic 
sequences provide strong information about transmission relationships, that is, 
they help to identify who infected whom, they contain no
information about the duration of infectiousness or the timing of recovery. As 
a result, parameters such as $\gamma$, which govern the epidemic time scale, 
remain weakly identifiable unless epidemiological information is also available.

These findings support the growing view that integrating mechanistic
epidemic models with genomic data provides a powerful framework for
statistical inference. At the same time, there are some limitations and
extensions that require further discussion:

First, all transmission graphs consistent with the observed data are treated as equally plausible. However, some transmission pathways may be biologically or epidemiologically more likely than others. A more realistic extension would assign graphs probabilities in $(0,1)$ reflecting their relative plausibility. Incorporating such weighted structures would provide a more nuanced representation of transmission uncertainty and may further improve parameter inference.

Second, we assumed observation at a complete
household-level data. Extensions to partially observed outbreaks,
irregular sampling times, or incomplete sequencing coverage would
increase realism and broaden applicability. Importantly, the
graph-resolved structure developed here provides a natural basis
for such extensions, as missing or uncertain components can be
treated within the same probabilistic framework.

Despite these limitations, the central methodological insight
remains robust: identifiability can be improved not only by
increasing temporal complexity, but by refining the structural
resolution of the model itself. By expanding the state space to
encode transmission graphs and constraining these graphs using
genetic information, we preserve the analytical advantages of
time-homogeneous household models while recovering partial
transmission-direction information.

This framework therefore opens the door to integrating genomic
resolution into structured epidemic models without abandoning
analytical tractability, and provides a principled foundation for
future work at the interface of transmission modelling and
pathogen genomics.

\section*{Acknowledgements}
Work supported by the Wellcome Trust (Grant Number 227438/Z/23/Z)

\section*{Code availability}
The code for this paper can be found at: 
\href{https://github.com/GSayyar/Improving-Parameter-Identifiability-in-Household-Transmission-Models-Using-Genomic-Data}{https://github.com/GSayyar/Improving-Parameter-Identifiability-in-Household-Transmission-Models-Using-Genomic-Data}.

\printbibliography

\appendix
	
\section{Final size and invariance to transmission order}\label{app1}

We explain here why making use of genomic information that resolves the true transmission route
does not allow for analysis using standard final-size formulas. \textcite{ludwig1975final} is usually credited with demonstrating that the final size distribution of an epidemic is invariant to the ordering of infection events in standard stochastic epidemic models. In the context of households \parencite{ball1997epidemics}, this means that the probability that each susceptible individual within a household escapes infection from outside the household can be expressed as $\exp(-\Lambda)$, where $\Lambda$ is the total force of external infection during the epidemic defined in \eqref{eqn:external-rate} above for our model, meaning that for our case
\begin{equation}
    \Lambda = \int_{t=1}^{\infty} \beta(t)\langle I(t) \rangle \, \mathrm{d}t.
\end{equation}
The time invariance assumption means that each member of the household of size $n$ can be assumed to avoid infection from outside with independent probability, meaning that the probability of $a$ household members being initially infected from outside is
\begin{equation}
    \Pr(a) = \binom{n}{a} (\exp(-\Lambda))^{(n-a)}(1-\exp(-\Lambda))^a , \qquad a\in\{0,\ldots,n\},
\end{equation}
and then the probability distribution for the number of further infections conditional on $a$ can be calculated through various methods with a review provided by \textcite{house2013big}.

As discussed by \textcite{pellis2008relationship}, however, standard final-size calculations do not assign probabilities to true transmission routes, which are instead summed over. For example, they might give the probability that 2 members of a size-3 household are ultimately infected, but would not attribute separate probabilities to the situations where this was due to two infections from outside versus where this was due to one infection from outside and an internal infection. As such, a final-size calculation cannot be used to evaluate our model likelihood \eqref{eqn:likelihood-sum}.

\section{Time-inhomogeneous household models}\label{app2}

For a single household the number of combined epidemiological states of all members is comparatively small, allowing us to formulate household models explicitly in terms of the  distribution of these combined states; this is in contrast to models of larger populations where the epidemiological state space is prohibitively large. This means that for infections where the assumption of memoryless transmission dynamics is appropriate, a within-household epidemic can be encoded as a continuous-time Markov chain on a finite state space. If we use $Q$ to denote the transition matrix of this Markov chain and  $P(t)$ to denote its state distribution at time $t$, then $P(t)$ satisfies a system of Kolmogorov forward equations of the form
	
\begin{equation}
\frac{\mathrm{d}P}{\mathrm{d}t} = QP,
\label{eq1}
\end{equation}\parencite{joe2022}. When $Q$ is constant in time, Equation~\eqref{eq1} has an explicit solution in terms of the matrix exponential,
\begin{equation}
	P(t) = \exp\left((t-t_0) Q\right)P(t_0),
\label{eq2}
\end{equation}
providing a mathematically elegant and tractable description of household epidemic dynamics exploited by e.g.\ \textcite{kinyanjui2018scabies} to obtain a significant speed-up.

As discussed above, in our case, we introduce explicit time dependence into the force of infection, meaning that 
the household state distribution satisfies the time-inhomogeneous linear set of equations
	\begin{equation}
		\frac{\mathrm{d}P}{\mathrm{d}t} = Q(t)P,
		\label{eq3}
	\end{equation}
where the transition matrix $Q(t)$ depends explicitly on time.
While this model is required for all but the simplest case of a piecewise constant external infection rate, which can be split into regions each obeying \eqref{eq1},
it introduces substantial mathematical and computational difficulties as the solution can no longer be written as a simple matrix exponential.

Given the analytical benefits of representation of the solution as a matrix exponentiall of the form \eqref{eq2}, it is natural to look for generalisations.
One formal representation of the solution to \eqref{eq3} is given by the Peano-Baker series \parencite{baake2011peano},
	$$
	P(t) = \left[I + \int_{t_0}^t Q(s_1)\,\mathrm{d}s_1 + 
    \int_{t_0}^t Q(s_1)\int_{t_0}^{s_1} Q(s_2)\,\mathrm{d}s_2\,\mathrm{d}s_1 + \cdots \right]P(t_0),
	$$
which involves an infinite series of nested integrals that are sometimes heuristically written as
\begin{equation}
	P(t) = \mathcal{T} \exp\left(\int_{t_0}^t Q(\tau)\,\mathrm{d}\tau
    \right)P(t_0),
\label{eq2}
\end{equation}
where $\mathcal{T}$ represents time ordering of the objects to its right. Although this series expansion is formally exact, when calculated for our system it rapidly becomes computationally impractical as the dimension of $Q(t)$ grows, which occurs quickly with increasing household size.
	
Another representation is given by the Magnus expansion \parencite{magnus1954exponential}, which expresses the solution as
	$$
	P(t) = \exp\big(\Omega(t)\big)P(t_0),
	$$
where $\Omega(t)$ is itself an infinite series involving time integrals and repeated commutators of $Q(t)$, for example
	\[
	\Omega(t) = \int_{t_0}^t Q(s)\,\mathrm{d}s + 
    \frac{1}{2}\int_{t_0}^t \mathrm{d}s_1 \int_{t_0}^{s_1} \mathrm{d}s_2 \,[Q(s_1),Q(s_2)] + \cdots.
	\]
In the absence of special commutation relationships, this expansion is also rarely tractable beyond very low-dimensional systems.

In realistic household models, where the state space grows combinatorially with household size, both the Peano-Baker series and the Magnus expansion are therefore typically usable only as numerical approximations rather than exact analytical tools. It is of course possible that more sophisticated analytical techniques could yield exact solutions in special cases; however, to the best of our knowledge, no general exact solution methods are currently available for time-inhomogeneous household models of realistic size.
We also performed preliminary investigation of Peano-Baker and Magnus expansions as numerical schemes, and believe that this is an interesting direction for future investigation, but beyond the scope of this study.
As a consequence, time-dependent household models, while providing a more realistic description of epidemic dynamics and
generally require local numerical approximation through time-stepping methods, which is the approach we have used here.

\end{document}